\documentclass[10pt,letterpaper]{article}
\usepackage[top=0.85in,left=2.75in,footskip=0.75in]{geometry}
\usepackage{booktabs,siunitx}

\usepackage{changepage}
\usepackage[utf8]{inputenc}
\usepackage{textcomp,marvosym}
\usepackage{graphicx,color}

\usepackage{fixltx2e}

\usepackage{amsmath,amssymb}

\usepackage{cite}

\usepackage{nameref,hyperref}

\usepackage[right]{lineno}

\usepackage{microtype}
\DisableLigatures[f]{encoding = *, family = * }

\usepackage{rotating}

\usepackage[aboveskip=1pt,labelfont=bf,labelsep=period,justification=raggedright,singlelinecheck=off]{caption}

\makeatletter
\renewcommand{\@biblabel}[1]{\quad#1.}

\def\d{\mathrm{d}}
\def\ds{\mathrm{ds}}
\def\e{\mathrm{e}}
\def\dt{\partial_t}

\def\nm{\mathrm{nm}}
\def\um{\mu\mathrm{m}}
\def\pN{\mathrm{pN}}
\def\nN{\mathrm{nN}}

\makeatother

\date{}

\begin{document}
\vspace*{0.35in}

\begin{flushleft}

{\Large
\textbf\newline{Dynamic curvature regulation accounts for the symmetric and asymmetric beats of   {\it Chlamydomonas}  flagella}
}
\newline\\
Pablo Sartori\textsuperscript{1,\Yinyang},
Veikko F. Geyer\textsuperscript{2,\Yinyang},
{ Andre} Scholich\textsuperscript{1},
Frank J\"ulicher\textsuperscript{1},
Jonathon Howard\textsuperscript{2,\dag}
\\
\bigskip
\bf{1} Max Planck Institute for the Physics of Complex Systems, Dresden, Germany
\\
\bf{2} Department of Molecular Biophysics and Biochemistry, Yale University, New Haven, Connecticut
\\

\bigskip

%
%
\Yinyang These authors contributed equally to this work.





\dag jonathon.howard@yale.edu

\end{flushleft}
\section*{Abstract}
Axonemal dyneins are the molecular motors responsible for the beating of cilia and flagella. These motors generate sliding forces between adjacent microtubule doublets within the axoneme, the motile cytoskeletal structure inside the flagellum. To create regular, oscillatory beating patterns, the activities of the axonemal  dyneins must be coordinated both spatially and temporally. It is thought that coordination is mediated by stresses or strains that build up within the moving axoneme, but it is not known which components of stress or strain are involved, nor how they feed back on the dyneins. To answer this question, we used isolated, reactivate axonemes of the unicellular alga  {\it Chlamydomonas} as a model system. We derived a  theory for beat regulation in a two-dimensional model of the axoneme. We then tested the theory by measuring the beat waveforms of wild type axonemes, which have asymmetric beats, and mutant axonemes, in which the beat is nearly symmetric, using high-precision spatial and temporal imaging. We found that regulation by sliding forces fails to account for the measured beat, due to the short lengths of {\it Chlamydomonas} cilia. We found that regulation by normal forces (which tend to separate adjacent doublets) cannot satisfactorily account for the symmetric waveforms of the \textit{mbo2} mutants. This is due to the model's failure to produce reciprocal inhibition across the axes of the symmetrically  beating axonemes. Finally, we show that regulation by curvature accords with the measurements. Unexpectedly, we found that the phase of the curvature feedback indicates that the dyneins are regulated by the dynamic (i.e. time-varying) component of axonemal curvature, but not by the static one. We conclude that a high-pass filtered curvature signal is a good candidate for the signal that feeds back to coordinate motor activity in the axoneme.

\section*{Author Summary}
The swimming of many microorganisms is powered by periodic bending motion of cilia and flagella. The flagellar beat results from a feedback: dynein motors generate sliding forces that bend the flagellum;  and bending leads to deformations and stresses, which feed back and regulate motors. Three alternative feedback mechanisms have been proposed: regulation by the sliding forces, regulation by the curvature of the axoneme, and regulation by the normal forces that tend to separate adjacent doublets. In this work we combine theoretical and experimental techniques to test whether any of these mechanisms can account for the waveforms of the short flagella of the unicellular alga {\it Chlamydomonas}. We show that the sliding control mechanism can not produce bend propagation for short flagella, which results in a poor fit to the data. Comparison of the waveforms of wild type {\it Chlamydomonas} with those of a mutant that has a symmetric beat argues against normal force regulation. By contrast, the  waveforms predicted by the curvature control model accord with the experimental data. Importantly, we make the surprising prediction that the motors respond to the time derivative of curvature, rather than curvature itself, hinting at an adaptive mechanism within the cilium. 


\section*{Introduction}
Cilia and flagella are long, thin organelles whose regular oscillatory bending waves propel cells through fluids and drive fluid flows across the surfaces of cells. The internal motile structure, the axoneme, contains nine doublet microtubules, a central pair of single microtubules, motor proteins in the axonemal dynein family and a large number of additional structural and regulatory proteins \cite{pazour_proteomic_2005}. The axonemal dyneins power the beat by generating sliding forces between adjacent doublets. The sliding is then converted to bending by constraints at the base of the axoneme (e.g. provided by the basal body) and/or along the length of the axoneme (e.g. nexin links) \cite{alberts_molecular_2002, brokaw_direct_1989,summers_adenosine_1971}.

While the constrained-sliding mechanism of bend formation is well established, it is not known how the activities of the dyneins are coordinated in space and time to produce the periodic beating pattern. It is thought that the 
beat is the result of feedback. The axonemal dyneins generate forces that deform the axoneme; the deformations, in turn, regulate the dyneins.
Because of the geometry of the axoneme, deformation leads to stresses and strains that have components in various directions (e.g. axial and radial). However, which component (or components) regulates the dyneins is not known.

\begin{figure}[htb]
\centerline{\includegraphics{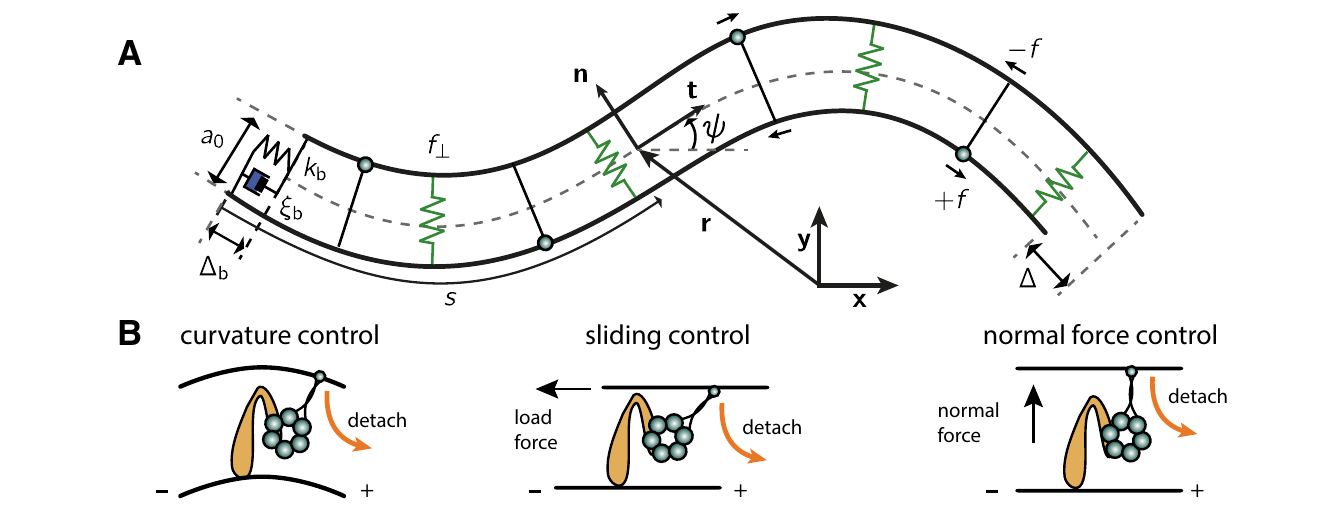}}
\caption{{\bf Opposing filaments model of the axoneme and dynein regulation.} ({\bf A}) Scheme of two opposing filaments bent by motors, as seen in the bending plane ${\bf x}{\bf y}$. The two filaments are constrained to have a spacing $a_0$. The dyneins step towards the base of the doublets. Dyneins sitting at the bottom filament with their head (blue circle) on the top filament produce a tensile force density $+f$ on the top filament, which tends to slide it towards the distal end; and a compressive force density $-f$ on the bottom filament. The dyneins sitting on the opposite filament create antagonistic forces. The local sliding displacement is given by $\Delta$, and the sliding at the base is $\Delta_{\rm b}$. The  spring and dashpot at the base correspond to the compliance of the base with stiffness $k_{\rm b}$ and friction coefficient $\xi_{\rm b}$. The green springs indicate the normal compliance which supports a normal force $f_\perp$. The position of the point at the arc length $s$ from the origin is ${\bf r}$, characterized by a tangent vector ${\bf t}$, a normal vector ${\bf n}$, and a tangent angle $\psi$ with respect to the horizontal axis of the lab-frame ${\bf xy}$. ({\bf B}) Schematic of dynein regulation mechanisms. Under {\it curvature control} the dynein head detaches due to an increase in curvature. In {\it sliding control} detachment is enhanced by a tangential loading force, and in {\it normal force control} it is the normal force that enhances detachment. Signs indicate doublets polarity.
\label{fig:axoscheme}}
\end{figure}

Three different, but not mutually exclusive, models for dynein regulation have been suggested in the literature, see Fig.~\ref{fig:axoscheme}. 
According to the sliding control model, dyneins are regulated by tangential forces acting parallel to the long axis of the microtubule doublets \cite{brokaw_molecular_1975, julicher_spontaneous_1997, camalet_generic_2000,riedelkruse_how_2007,morita2004effects}. 
According to the curvature control model, dyneins are regulated by doublet curvature \cite{brokaw_bend_1971, brokaw_thinking_2009, machin_wave_1958}. 
According to the normal force control model, dyneins are regulated by transverse forces that act to separate adjacent doublets when they are curved \cite{lindemann_model_1994,brokaw2009thinking}. Which of these mechanisms regulates the beat of the axoneme is not known.

In this work, we isolated {\it Chlamydomonas} axonemes, and tracked them with high spatial and temporal resolution. Performing a Fourier decomposition of the data, we observed that the beat of {wild type} and {\it mbo2} mutant axonemes have similar dynamics. The primary difference lies in the static asymmetry characteristic of the {wild type} beat, which  is lacking in the mutant \cite{geyersub}. By developing a two-dimensional theory for the axonemal beat in the presence of a static asymmetry, we show that sliding control, curvature control, and normal force control can all, in principle, give raise to wave propagation in asymmetric axonemes. However, the observed beats of {wild type} cells and {\it mbo2} mutants are not consistent with sliding and normal force control. By contrast, curvature control accorded with the {wild type} and {\it mbo2} beats provided that the curvature signal adapts (or is insensitive to) the static component of curvature.

\section*{Theoretical model}
\subsection*{Periodic beat of an asymmetric cilium}
We model the axonemes imaged in our experiments as two opposing inextensible filaments immersed in a  fluid (see Fig.~\ref{fig:axoscheme}). In this two-dimensional model, the filaments are connected by motor proteins, which are oriented in both directions and so can slide filaments in either direction. This captures the crucial idea that motors on opposite sides of the axoneme generate antagonistic bending forces \cite{satir1989splitting}. Elastic elements keep the filaments at a constant distance $a_0$ from each other. The position and shape of the filament pair is given at each time by the vector ${\bf r}(s)$, a function of the arc length $s$ along the centerline between the filaments. Calculating the tangent vector as ${\bf t}(s)=\dot{{\bf r}}(s)$, where dots denote arc length derivatives, allows us to define the tangent angle $\psi(s)$ with respect to the horizontal axis of the lab-frame that characterizes the shape of the filament at each time. For a given filament shape, the pair of filaments will have a local sliding displacement $\Delta(s)$ with respect to the other \cite{everaers1995fluctuations, camalet2000generic, brokaw1971bend}. The sliding is related to the shape of the axoneme via
\begin{align}
\Delta(s)=\Delta_{\rm b}+a_0(\psi(s)-\psi(0))\quad,\label{eq:sliding}
\end{align}
where $\Delta_{\rm b}$ is the basal sliding { and $a_0$ is the spacing between the filaments} (see Fig.~\ref{fig:axoscheme}).

{ To calculate how the shape of the axoneme depends on the active motor forces $f$, as well as the passive mechanical properties of the axoneme (such as its bending rigidity { $\kappa$}) and the viscous properties of the fluid, we establish a force balance, see {\it Appendix}. To calculate the mechanical forces that the axoneme exerts on the fluid we introduce a work functional $U$ that contains the effect of axonemal rigidity and active motors, and differentiate with respect to {\bf r}. This force is balanced by the friction force of the fluid, proportional to the velocity of the axoneme at each point, and so we have  $-{\bf\hat{\Pi}}\cdotp\partial_t{\bf r}=\delta U/\delta{\bf r}$, where  ${\bf\hat{\Pi}}$ is the friction matrix  \cite{lauga2009hydrodynamics}.} For a slender body at low Reynolds number we have ${\bf\hat{\Pi}}=\xi_{\rm n}{\bf nn}+\xi_{\rm t}{\bf tt}$, with $\xi_{\rm n}$ and  $\xi_{\rm t}$ the normal and tangential hydrodynamic friction coefficients of the cilium and ${\bf n}(s)$ a unit vector normal to ${\bf t}(s)$ (see Fig.~\ref{fig:axoscheme}). Force balance provides non-linear equations of motion which are given in the {\it Appendix}.  We then look for periodic solutions of these equations at the beat frequency of the axoneme. 

For observed periodic beats, we can decompose the tangent angle into Fourier modes
\begin{align}
\psi(s,t) =\sum_{n=-\infty}^{\infty}\psi_n(s)\exp(i n \omega t)\quad ,\label{eq:modes}
\end{align}
where $\omega$ is the fundamental angular frequency of the beat ($\omega/2\pi$ is thus the beat frequency), $t$ is time, and $\psi_n$ are the harmonics which satisfy $\psi_{-n}=\psi_n^*$ to keep the angle real. We refer to $n=0$ as the static mode and to $n=1$ as the fundamental mode.  The same decomposition can be done for the sliding force $f$ and the tension $\tau$ along the cilium.  For each value of $n$, we obtain an equation of motion. The $n=0$ equation allows us to calculate the time-averaged shape. The $n=1$ equation allows us to calculate the motion at the beat frequency.

The time-averaged shape is calculated from the static force balance $a_0 F_0=\kappa\dot{\psi}_0$, where $F_0$ is the zeroth mode of the integrated sliding force $F=-\int_s^L f(s')\d s'$. As we will show later, {\it Chlamydomonas} axonemes have a static curvature $C_0$ that is  approximately constant along the arc length, $\dot{\psi}_0(s)=C_0$ \cite{geyersub,eshel1987new}.  According to the static force balance, bending the axoneme with a constant curvature requires accumulation of forces at the distal end \cite{mukundan2014motor}. The static component of the sliding force density is thus
\begin{align}
f_0=-\delta(s-L)\kappa C_0/a_0\quad,\label{eq:statfor}
\end{align}
where $\delta$ is the Dirac delta function{, and the minus sign comes from the fact that dynein is a minus end directed motor which has a negative sliding velocity. See {\it Appendix} for details on the sign convention.}

To obtain the dynamic motion at the beat frequency, we expand the non-linear dynamic equations on the dynamic modes. Keeping the linear term we obtain the following equation of motion for the fundamental mode ($n=1$)
\begin{align}
i\bar{\omega}\psi_1&=-\ddddot{\psi}_1 + \ddot{f}_1+(1+\bar{\zeta})\bar{C}_0 \dot{\tau}_1 +\bar{\zeta} \bar{C}_0^2(\ddot{\psi}_1 - f_1)\nonumber\\
\bar{\zeta}\ddot{\tau}_1 - \bar{C}_0^2\tau_1&=-(1+\bar{\zeta})\bar{C}_0(\dddot{\psi}_1- \dot{f}_1)\quad,
\label{eq:asym}
\end{align}
which has constant coefficients and has to be supplemented by appropriate boundary conditions. Here all quantities have been made dimensionless as indicated in {\it Appendix}. There are three dimensionless constants noted with overbars: the dimensionless static curvature $\bar{C}_0=C_0L$; the normalized friction $\bar{\zeta}=\xi_{\rm n}/\xi_{\rm t}$; and the normalized frequency $\bar{\omega}=\omega\xi_{\rm n}L^4/\kappa$, which is sometimes referred  to as the ``sperm number'' \cite{lauga2009hydrodynamics}. Equation~\ref{eq:asym} shows how an oscillatory active sliding force $f_1$ produces dynamic bending of the cilium under appropriate parameter and boundary conditions. 

Equation  \ref{eq:asym}  is the equation of motion for an asymmetrically beating axoneme. It generalizes the equation for a symmetric axoneme \cite{machin_wave_1958,riedelkruse_how_2007,julicher_spontaneous_1997}, for which the static curvature is $C_0=0$ and the axial tension $\tau_1=0$.  For $C_0\neq0$ new terms appear, and the system of equations is of order six rather than four in the symmetric case. The magnitude of the new terms can be estimated by considering the plane wave limit, in which $\psi_1=\exp(-2\pi is/\lambda)$, with $\lambda$ the wavelength. For example, the fourth term in the right hand side of the equation is of order $\bar{\zeta} \bar{C}_0^2(2\pi)^2/\lambda^2$ and is in phase with the first term, that is also present in the symmetric theory and is of order $(2\pi)^4/\lambda^4$. For {\it Chlamydomonas} axonemes  $\lambda\sim L$ and $C_0\sim\pi/L$, and since $\zeta\approx2$ \cite{johnson1979flagellar, friedrich2010high} the ratio of these terms is $\sim0.5$, thus the new contributions due to the asymmetry can not be neglected. A similar reasoning shows that the tension term is in anti-phase, and its contribution is of order $\sim1$. This shows that the additional terms are non-neglibible. Thus, for the large observed asymmetry of the {\it Chlamydomonas} axoneme, we expect that in general coupling between the $n=0$ and $n=1$ modes  significantly modifies the dynamics of the beating axoneme. Surprisingly, we will later show that this is not the case for beats regulated by curvature.

\subsection*{Three mechanisms of motor control}
Equation~\ref{eq:asym} shows how an oscillating sliding force can produce a dynamic beating pattern. However, we do not expect the motor proteins themselves to be oscillators that drive axonemal motion at the beat frequency. Rather, we expect that the sliding forces generated by the dyneins are regulated directly or indirectly by the shape of the axoneme through feedback. In this view, the motor force produces a bend, and the  bend regulates the motor forces. Here we consider three different types of feedback motor regulation: sliding control, in which motors are regulated by sliding of the filaments $\Delta$; curvature control, in which they are regulated by the curvature $\dot{\psi}$; and normal force control, in which motors are regulated by the normal force $f_\perp$ that keeps the filaments spacing  at $a_0$, see  Fig.~\ref{fig:axoscheme}.

The sliding, curvature, and normal force can be decomposed in Fourier modes just as the angle $\psi$.
The fundamental mode of the sliding, $\Delta_1$, relates to that of the angle, $\psi_1$, and the basal sliding, $\Delta_{{\rm b},1}$, through Eq.~\ref{eq:sliding}, which is linear. Thus in sliding and curvature control the feedback is linear. In the case of the normal force control, however, the feedback from shape is non-linear, see {\it Appendix}, and the fundamental mode is given by
\begin{align}
f_{\perp,1} =C_0(F_1+ \kappa \dot{\psi}_1/a)\quad\quad,
\label{eq:fndyn}
\end{align}
where $F_1$ is the fundamental mode of the integrated sliding force. This expression vanishes for the symmetric case in which $\dot{\psi}_0=0$, indicating that in symmetric beats  the normal force only has higher harmonics. Thus the asymmetry $C_0\neq0$ of the {\it Chlamydomonas} cilium opens a way to regulation by normal forces, something impossible in symmetrically beating cilia of which {bull sperm} is an approximate example \cite{riedelkruse_how_2007}.

The most general expression for the dependence of the sliding force on sliding, curvature and normal force is
\begin{align}
f_{ n}=\chi(n \omega)\Delta_n + \beta(n \omega)\dot{\psi}_n + \gamma(n \omega)f_{\perp,n}\quad ,
\label{eq:fullmotmod}
\end{align}
where $n>0$ is a mode-index, and $\chi(\omega)$, $\beta(\omega)$ and $\gamma(\omega)$ are complex frequency dependent coefficients describing the response to sliding, curvature and normal forces, respectively. { These are the generalization to active motors of the familiar linear response of passive systems. For example, the basal force of the axoneme is given by  $F_{\rm b}=k_{\rm b}\Delta_{\rm b}+\xi_{\rm b}\partial_t\Delta_{\rm b}$, and thus the fundamental mode of the basal force is given by $F_{{\rm b},1}=\chi_{\rm b}(\omega)\Delta_{{\rm b},1}$ with $\chi_{\rm b}(\omega)=k_{\rm b}+i\omega\xi_{\rm b}$. Similarly, the coefficients $\chi$, $\beta$ and $\gamma$ will depend on active motor properties.} Equation \ref{eq:fullmotmod} expresses how changes in the shape of the axoneme affect the motor force, while Eq.~\ref{eq:asym} shows how the motor force influences the axonemal shape. Together, they form a dynamical system which can become unstable and produce spontaneous oscillations \cite{julicher_spontaneous_1997,camalet1999self}. At the critical point these oscillations are perfectly periodic and we only retain the fundamental mode from Eq.~\ref{eq:fullmotmod}. Finally, we note that for $n=0$ the response coefficients must vanish, because the observed constant static curvature implies that the static component of the force vanishes all along the length except at the very distal end of the axoneme (see Eq.~\ref{eq:statfor}). While for now we will take this for granted, we will later demonstrate that this is indeed the case for beats resulting from dynamic curvature control. 

\begin{figure}[htb]
\begin{adjustwidth}{-2.25in}{0in}
{\includegraphics{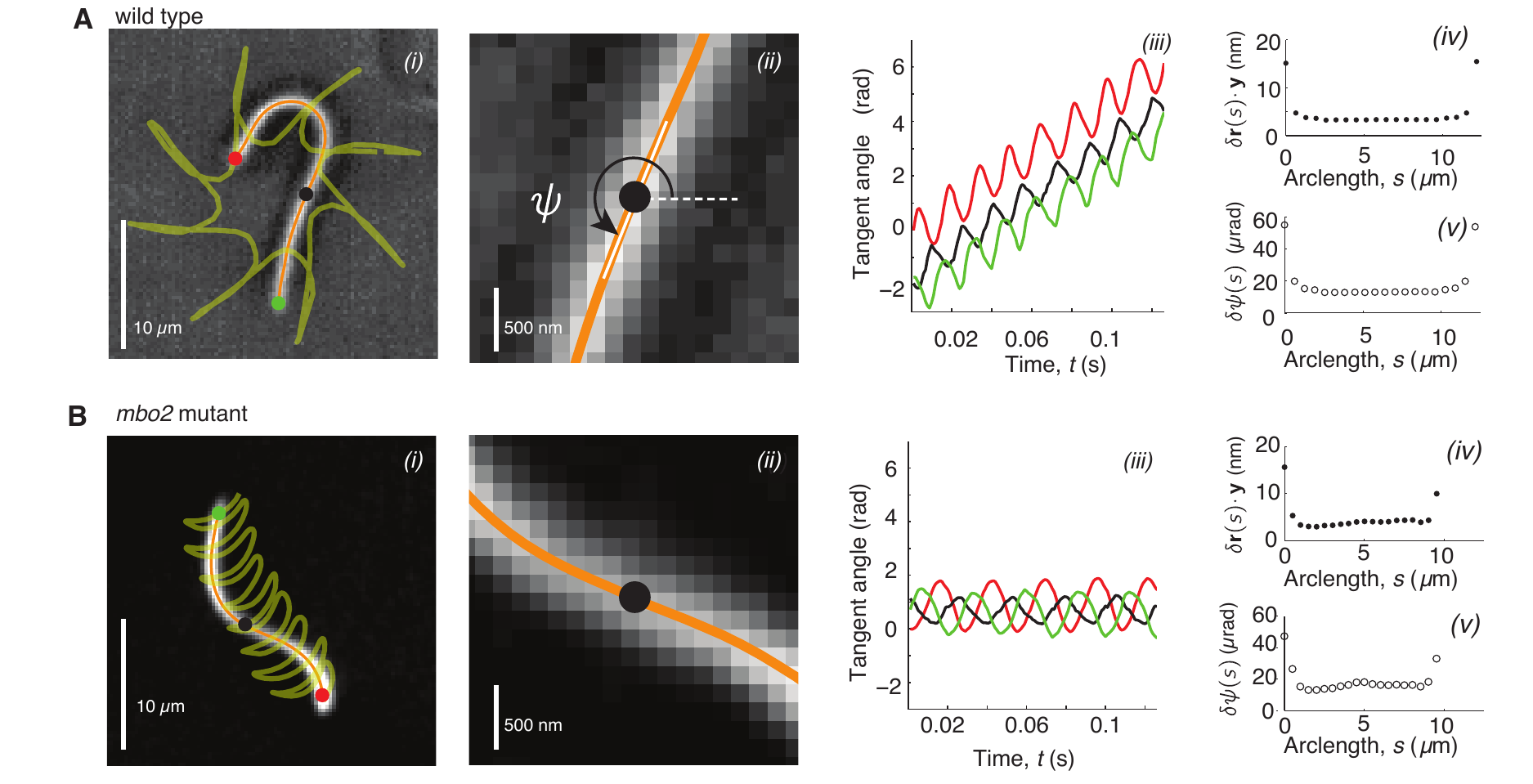}}
\caption{{\bf High precision tracking of isolated axonemes}  Panel A corresponds to {\it Chlamydomonas} wild type axonenes, panel B to \textit{mbo2} mutant axonemes. {({\bf A})} {(\textit{i})} Inverted phase-contrast image of a wild type axoneme. The orange curve represents the tracked centerline. The points depict the basal end (red), the distal end (green) and the center position (black) of the axoneme. The yellow line depicts the trajectory of the basal end. {(\textit{ii})} Same image as in A\textit{i}, magnified around the center region. The tangent angle $\psi(s,t)$ is defined with respect to the lab frame. {(\textit{iii})} Tangent angle  of three different arc length positions (depicted in A\textit{i})  as a function of time. The linearly increasing tangent angle corresponds to the rotation of the axoneme during swimming. Mean uncertainty for 1000 adjacent frames of the tracked shape $\delta{\bf r}\cdotp{\bf y}$ in {(\textit{iv})} (the $\delta{\bf r}\cdotp{\bf x}$ error gives the same result) and tangent angle $\delta \psi$ in {(\textit{v})}. {({\bf B})} is analogous to (A) but for {\it mbo2} mutant axonemes.}
\label{fig:tracking}
\end{adjustwidth}
\end{figure}

\section*{Experimental procedure}
\subsection*{Preparation, reactivation and imaging of axonemes}
	In this study, axonemes from \textit{Chlamydomonas reinhardtii} wild type cells (CC-125 wild type mt+ 137c, 
 R.P. Levine via N.W. Gillham, 1968) and mutant cells that {m}ove {b}ackwards {o}nly (CC-2377 \textit{mbo2} mt-, David Luck, Rockefeller University, May 1989) were purified and reactivated. The procedures described in the following are detailed in \cite{alper2012reconstitution}.
 
Chemicals were purchased from Sigma Aldrich, MO if not stated otherwise. In brief, cells were grown in TAP+P medium under conditions of illumination (2x75 W, fluorescent bulb) and air bubbling at $24\,^{\circ}\mathrm{C}$ over the course of 2 days, to a final density of $10^6\,{\rm cells/ml }$. Flagella were isolated using dibucaine, then purified on a 25$\%$ sucrose cushion and demembranated in HMDEK (30 mM HEPES-KOH, $5\,{\rm mM}$ MgSO$_{\rm 4}$, $1\,{\rm mM}$ DTT, $1\,{\rm mM}$ EGTA, $50\,{\rm mM}$ potassium acetate, pH 7.4) augmented with $1\%$ (v/v) Igpal and $0.2\,{\rm mM}$ Pefabloc SC.
 The membrane-free axonemes were resuspended in HMDEK plus $1\%$ (w/v) polyethylene glycol (molecular weight 20 kDa),  $30\%$ sucrose, $0.2\,{\rm mM}$ Pefabloc and stored at $-80\,^{\circ}\mathrm{C}$. Prior to reactivation, axonemes were thawed at room temperature, then kept on ice. Thawed axonemes were used for up to 2\,hours.
 
 Reactivation was performed in flow chambers of depth $100\,\mu {\rm m}$, built from easy-cleaned glass and double-sided sticky tape. Thawed axonemes were diluted in HMDEKP reactivation buffer containing $1\,{\rm mM\, ATP}$ and a ATP-regeneration system (5 units/ ml creatine kinase, 6\,mM creatine phosphate) used to maintain the ATP concentration. The axoneme dilution was infused into a glass chamber, that was blocked using casein-solution (solution of casein from bovine milk, 2 mg/mL, for 10 minutes) and then sealed with vacuum grease. Prior to imaging, the sample was equilibrated on the microscope for 5 minutes and data was collected for a maximum time of 20 minutes.
 
The reactivated axonemes were imaged by either phase constrast microscopy  (wild type axonemes) or darkfield microscopy (\textit{mbo2} axonemes).  Phase contrast microscopy was set up on an inverted Zeiss Axiovert S100-TV microscope using a Zeiss $63\times$ Plan-Apochromat NA 1.4 Phase3 oil lens in combination with a $1.6\times$ tube lens and a Zeiss oil condenser (NA 1.4). Data was acquired using a EoSens 3CL Cmos highspeed camera. The effective pixel size was 139\,nm/pixel.
Darkfield microscopy was set up on an inverted Zeiss Axiovert 200 microscope using a Zeiss $100\times$ Plan-Neofluar NA iris 0.7-1.4 oil lens in combination with an $1.2\times$ tube lens and a Zeiss oil Darkfield (NA 1.4). Data was acquired using a pco dmaxS highspeed camera.
In both cases the illumination was performed using a Sola light engine with a 455 LP filter. Movies of up to 3000 frames were recorded at a frame rate of 1000 fps. The sample temperature was kept constant at
$24^{\circ}{\rm C}$ using an objective heater (Chromaphor).

\subsection*{High precision tracking of isolated axonemes} 
To localize the position of the axoneme shapes in each frame of the recorded movie with nm precision, the Matlab based software tool FIESTA was used \cite{Ruhnow20112820}. Prior to tracking, movies were background subtracted to remove static inhomogeneities arising from uneven illumination and dirt particles. The background image contained the mean intensity in each pixel calculated over the entire movie. This procedure increased the signal to noise ratio by a factor of 3. Phase contrast images were then inverted; darkfield images were tracked directly. 

The tracking algorithm FIESTA uses manual thresholding to determine the filament skeleton, which is then divided into square segments. During tracking, the filament position in each segment is determined independently. For tracking, a segment size of 733 nm (approximately 5x5 pixels) was used, corresponding to the following program settings: a full width at half maximum of $750$\,nm, and a ``reduced box size for tracking especially curved filaments''  of 30 \%.  Along the arc length of each filament, 20 equally distributed segments were fitted using two dimensional Gaussian functions. Two  examples of spline fitted shapes  are presented in Figure \ref{fig:tracking}\,A\textit{i} and B\textit{i} superimposed on the tracked image. The mean localization uncertainty of the center position of each of these segments was about $5\,{\rm nm}$ (see Figure \ref{fig:tracking}~A\textit{iv} and B\textit{iv}). For localization of the ends, the program uses a different fitting function, resulting in an increased uncertainty.

\section*{Results}

\subsection*{Quantification of the beat of wild type and \textit{mbo2} cilia}
To test the different mechanisms of beat regulation we performed high precision measurements of the flagellar waveform, see {\it Experimental Procedures} and Fig.~\ref{fig:tracking}. Using the tracking software developed in \cite{Ruhnow20112820} we calculated the temporal trajectories of 20 points along the arc length of the axoneme. The uncertainty of the shape in the ${\bf xy}$ space was $\sim5\,{\rm nm}$ and the uncertainty in the tangent angle was  $~20\,\mu{\rm rad}$ (see panels $iv$ and $v$ in Fig.~\ref{fig:tracking}). The latter corresponds to a sliding displacement between adjacent doublet microtubules of only $0.06\,{\rm nm}$.


%

\begin{figure}[!htb]
\centerline{\includegraphics{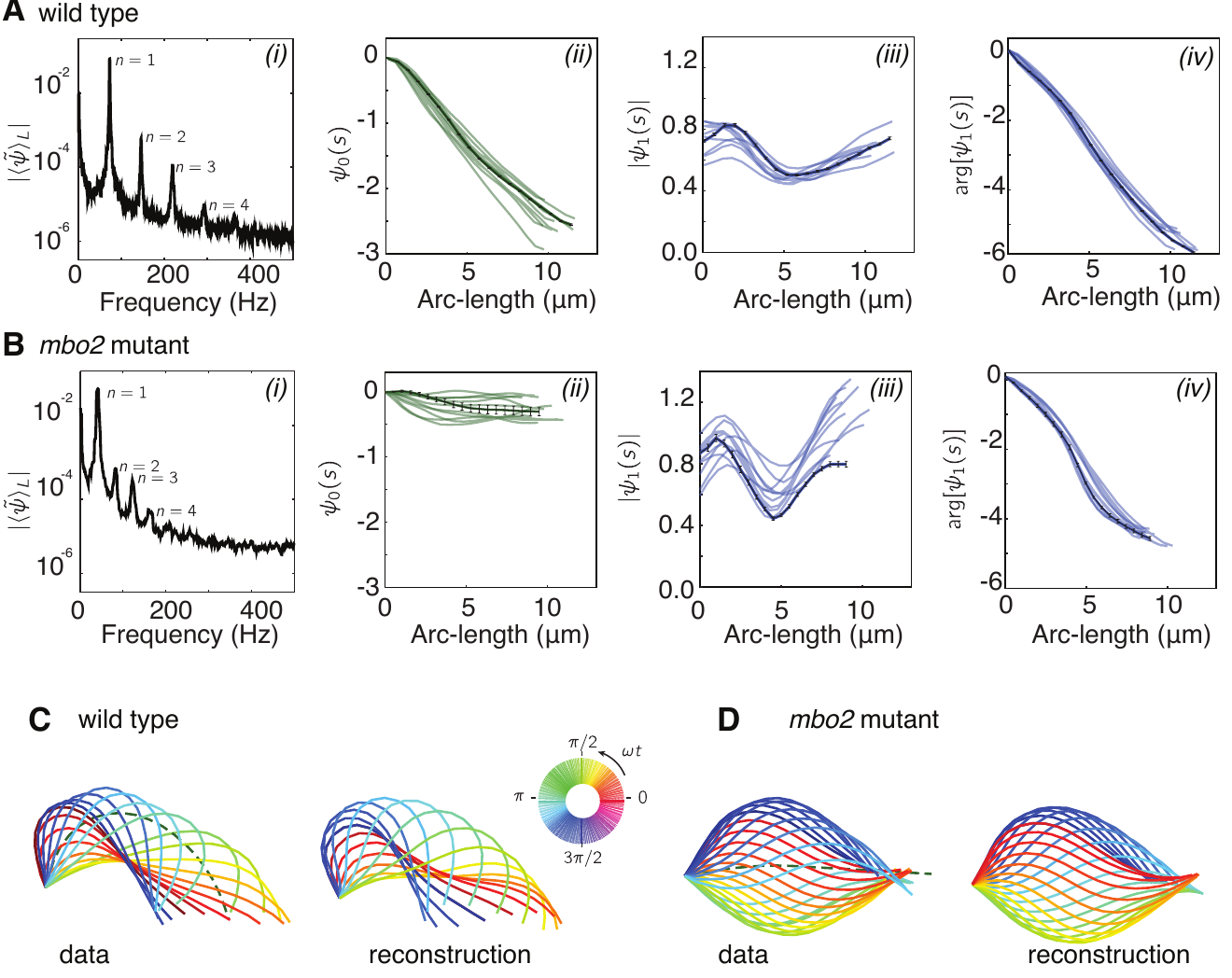}}
\caption{{\bf Fourier decomposition of the beat.} Panel A shows the Fourier decomposition of the waveform of wild type axonemes, panel B the decomposition of the \textit{mbo2} axoneme waveform. {({\bf A})} {(\textit{i})} Power spectrum of the tangent angle averaged over arc length. The fundamental mode ($n=1$)  and three higher harmonics ($n=2,3,4$) are labeled. {(\textit{ii})} Angular representation of the static ($n=0$) mode as a function of arc length. The constant slope indicates an arc length independent static curvature $\dot{\psi}_0=C_0$. {(\textit{iii})} The amplitude and phase (argument) of the fundamental mode are shown in {\textit{iii}} and {\textit{iv}}, respectively. The linear decrease in phase indicates steady wave propagation. The data of a representative axoneme is highlighted in the panels \textit{ii}--{\it iv}, with error bars indicating the standard error of the mean calculated by hexadecimation. {({\bf B})} Equivalent plots to (A) for \textit{mbo2} axonemes.  {({\bf C})} Beat shapes of one representative beat cycle of the wild type axoneme highlighted in panel A (left panel, data) and shapes reconstructed from the superposition of the static and fundamental modes, neglecting all higher harmonics. The progression of shapes through the beat cycle is represented by the rainbow color code (see inset). {({\bf D})} Same as (C) for an \textit{mbo2} mutant axoneme.  
\label{fig:tracked}}
\end{figure}

Because the beat of {\it Chlamydomonas} is periodic in time, it is convenient to decompose the tangent angle $\psi(s, t)$ into Fourier modes $\psi_n$, see Eq.~\ref{eq:modes}.  Before doing so, we note that wild type {\it Chlamydomonas} axonemes swim in circles at a slow angular rotation speed { $\omega_{\rm rot}\sim30\,{\rm rad}/{\rm s}$}, see $i$ and $iii$ in Fig.~\ref{fig:tracking}A. While the effect of this rotation is small for a single beat it becomes large for a long time series. Before performing the Fourier decomposition we subtracted $\omega_{\rm rot}t$ from the tangent angle $\psi(s, t)$ \footnote{For simplicity we use the same notation for $\psi(s, t)$ and $\psi(s, t)-\omega_{\rm rot}t$.}. The power spectrum of the tangent angle (averaged over the flagellar length) shows clear peaks at harmonics of its fundamental frequency, see Fig.~\ref{fig:tracked}A\textit{i}. Note that the peak of the fundamental mode $n=1$ accounts for $90\%$ of the total power spectrum,  and so we neglect the higher harmonics $n=2,3,4,\ldots$ for reconstructing the flagellar shape.


The amplitude of the static mode ($n=0$) and the amplitude and phase of the fundamental mode ($n=1$) are shown in Fig.~\ref{fig:tracked}\textit{ii}--\textit{iv}. The main difference between the wild type and mutant axonemes comes from the static angular mode $\psi_0$. For wild type axonemes, $\psi_0$ decreases with an approximately constant slope over arc length. This corresponds to a constant static curvature $\sim0.25\,{\rm rad}/\um$, and indicates that the static mode of the shape is a circular arc of radius $\sim4\,\mu{\rm m}$. The static curvature of wild type axonemes leads to a highly asymmetric waveform. In contrast, {\it mbo2} mutant axonemes have a very small static mode with a curvature $\sim0.025\,{\rm rad}/\um$, and the resulting beat is approximately symmetric. In comparison to the large difference in the static mode between wild type and mutant axonemes, the fundamental modes are similar. The amplitude of $\psi_1$ is roughly constant and has a characteristic dip in the middle. The argument of $\psi_1$, which determines the phase profile of the wave, decreases at a roughly constant rate in both cases. Since the total decay is about $-2\pi$, the wavelength of the beat is approximately equal to the length of the axoneme. In summary, both wild type and mutant cells have an approximately sinusoidal dynamic beat whose amplitudes drop in the middle of the axoneme and whose phases decrease monotonically, consistent with a beat wavelength of about $13\,\mu{\rm m}$.


\subsection*{Motor regulation in  the axoneme: experiments vs theory}
The response of the motor force to strains and stresses of the axoneme described by Eq.~\ref{eq:fullmotmod}  allows for regulation via sliding, curvature and normal forces. While it is possible that all three mechanisms of motor control are involved in regulation of the axonemal beat, we now show each individual mechanism alone is capable of producing dynamic bending patterns.

\begin{figure}[htb]
\centering
\includegraphics{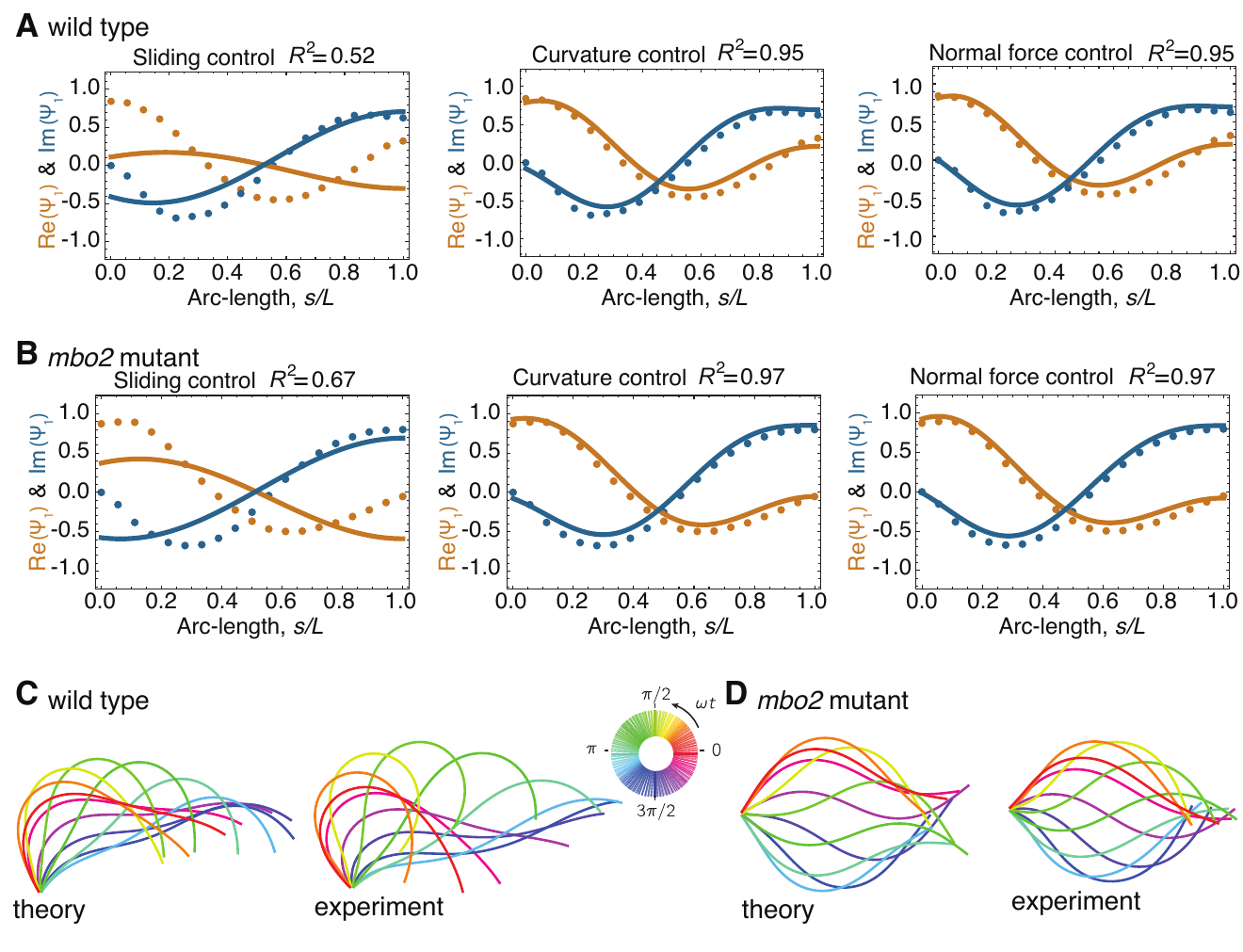}
\caption{
{\bf Comparison of theoretical and experimental  beating patterns}.  ({\bf A}) Comparison of the theoretical (lines) and experimental (dots) beating patterns of a typical wild type axoneme. The real and imaginary part of the first mode of the tangent angle $\psi_1(s)$ is plotted for beats resulting from sliding control, curvature control, and normal force control. ({\bf B}) Analogous to (A)  for {\it mbo2}. Note that here also curvature control and normal force control provide good agreement, but not sliding control. ({\bf C}) and ({\bf D}) Theoretical and experimental shape reconstruction in position space for the wild type and {\it mbo2} beats under curvature control. \label{fig:fits} }
\end{figure}

Because the axonemal beat is dominated by its static mode ($n=0$) and fundamental   dynamic mode ($n=1$), see Fig.~\ref{fig:tracked}, we constrain our physical model to one dynamic mode, which corresponds to the critical point of an oscillatory Hopf bifurcation \cite{camalet1999self,camalet_generic_2000}. Within our theoretical description, {\it sliding control} corresponds to the case in which $\beta=0$ and $\gamma=0$, and the motor force only responds to sliding changes through $\chi(\omega)=\chi'+i\chi''$, with single and double primes denoting real and imaginary parts respectively and $i$ the imaginary unit. Additionally, because the active response must dominate for oscillations to occur, we have $\chi'<0$ and $\chi''<0$ \cite{julicher_spontaneous_1997, camalet1999self}.  In {\it curvature control}, the motors have a passive response to sliding (corresponding to the slope of their force-velocity curves), so that $\chi'>0$; the motors are actively regulated by curvature, thus $\beta''<0$; and they are not regulated by normal force, that is $\gamma=0$. Finally, in {\it normal force control} there is a passive response to sliding, $\chi'>0$, an active response to normal forces, with $\gamma'>0$ and  $\gamma''<0$, and no response to curvature, $\beta=0$. \footnote{For backward traveling waves the signs of $\beta',\beta''$ and $\gamma',\gamma''$ change, but not those of $\chi',\chi''$.}

\begin{table}[!ht]
\begin{adjustwidth}{-2.25in}{0in}
\caption{\textbf{Parameters for beat generation in wild type axonemes.} }
\begin{tabular}{lllll}
\toprule
 & Sliding control  & Curvature control & Normal force control \\
\toprule
$R^2\;(\%)$ & $49\pm4$ &$95\pm1$&$95\pm1$\\
\midrule
$\chi',\chi''\; (\nN\cdotp\um^{-2})$ &$-14.7\pm3.6,-1.31\pm0.16$&
$ 24\pm4,0 $&
$16.0\pm4.4,0$\\
\midrule
$\beta',\beta''\;(\nN)$&$0$&$0,-7.1\pm0.5$&$0$\\
\midrule
$\gamma',\gamma''$&$0$&$0$&$0.123\pm0.052,1.96\pm0.21$\\
\midrule
$\chi'_{\rm b},\chi''_{\rm b}\; (\nN\cdotp\um^{-1})$ &$51.3\pm 0.9,3.35\cdotp10^3\pm 7.16\cdotp10^3$&
$51\pm15,2.6\pm0.9$&
$96.3\pm139.0,15.2\cdotp10^{3}\pm40.9\cdotp10^{3}$\\
\midrule
$\Delta_{{\rm b},0}\;(\nm)$&$-37\pm37$&$-33\pm11$&$-131\pm178$\\
\midrule
$|\Delta_{{\rm b},1}|\;(\nm)$&$21\pm17$&$19.5\pm4.7$&$10.7\pm9.7$\\
\bottomrule
\end{tabular}
\label{tab:wtfits}
\begin{flushleft} 
Curvature control and normal force control result in high $R^2$ values. Sliding control, unable to produce bend propagation (see Fig.~\ref{fig:fits}), produces a low $R^2$. The values reported are mean and { standard deviation} calculated with $9$ axonemes (when the standard deviation was larger than the mean it was replaced by the mean itself). The average static curvature is $C_0=-0.232\pm\,0.009\um^{-1}$. Note that, for curvature control beats, using the value of $\beta$ here reported and an estimate curvature of $0.1\,{\rm rad}/\um$ results in a sliding force of $\sim700\,\pN/\um$. Since the motor density of an active half of the axoneme is $~\sim 500\,\um^{-1}$ the individual motor force is about $~\sim1\,\pN$. For the case of normal force control the sliding force generated by the motors is of the order of the normal force that they experience, since $|\gamma|\approx 2$.
\label{tab:wtfits}
\end{flushleft} 
\end{adjustwidth}
\end{table}

For wild type cells, we compared the modes obtained from the three motor models with the observed beats (see {\it Appendix} for details on fitting procedure and parameters used). The result of a typical fit is shown in Fig.~\ref{fig:fits}A, where the only two free parameters were the basal stiffness $k_{\rm b}$ and viscosity $\xi_{\rm b}$.  As we can see, the predicted real and imaginary part of $\psi_1(s)$ agree well with the data for a wild type axoneme in the cases of curvature control and normal-force control, but not for sliding control. In fact, in the beats predicted by sliding control, the real and imaginary parts are in anti-phase. This results in a standing wave with no bend propagation.

\begin{table}[!ht]
\begin{adjustwidth}{-2.25in}{0in}
\caption{\textbf{Parameters for beat generation in  \textit{mbo2} mutant axonemes.} }
\begin{tabular}{lllll}
\toprule
 & Sliding control & Curvature control & Normal force control\\
\toprule
$R^2\;(\%)$ & $72\pm5$ &$95\pm1$&$96\pm1$\\
\midrule
$\chi',\chi''\; (\nN\cdotp\um^{-2})$ &$-18.2\pm1.2,-0.52\pm0.12$&
$ 21\pm3,0 $&
$12.7\pm5.2,0$\\
\midrule
$\beta',\beta''\;(\pN)$&$0$&$0,-7.1\pm0.5$&$0$\\
\midrule
$\gamma',\gamma''$&$0$&$0$&$1.52\pm1.52,32\pm25$\\
\midrule
$\chi'_{\rm b},\chi''_{\rm b}\; (\nN\cdotp\um^{-1})$ &$37\pm 17,1.98\pm 0.87$&
$2.5\pm0.2,2.8\pm2.1$&
$2.8\pm0.3,16.4\pm7.2$\\
\midrule
$\Delta_{{\rm b},0}\;(\nm)$&$-5.7\pm5.1$&
$-66\pm40$&
$-58\pm36$\\
\midrule
$|\Delta_{{\rm b},1}|\;(\nm)$&$46\pm9$&
$54\pm10$&
$69\pm25$\\
\midrule
\end{tabular}
\label{tab:mbo2fits}
\begin{flushleft}
As for wild type beats curvature control and normal force control provide very good fits, and sliding control does not. Values indicated are averages and { standard deviations} for $9$ axonemes (when the standard deviation was larger than the mean it was replaced by the mean itself). The static curvature was $C_0=-0.0276\pm\,0.005\um^{-1}$.  Note that the values of $\gamma$  in normal force control are very spread and different relative to those obtained for wild type fits. In fact, in one case we obtained $\gamma\approx80$, indicating that motors must amplify the normal force they sense by almost two orders of magnitude. The values for curvature control are very similar to those of wild type fits.
\label{tab:mbo2fits}.
\end{flushleft}
\end{adjustwidth}
\end{table}

The beating pattern obtained by curvature control is compared to the experimental reconstruction in Fig.~\ref{fig:fits}C. The good agreement reinforces the conclusion from Figure 4A that the curvature control model accords with the experimental data for wild type cells. Similar good agreement for wild type cells was found with the normal-force model. Table \ref{tab:wtfits} summarizes average parameters resulting from the fits of 9 different axonemes.


We compared theory and experiments for the symmetric beat of the mutant \textit{mbo2}, where the static curvature is reduced by at least one order of magnitude compared to wild type beats, see Fig.~\ref{fig:tracked}. In this case the results  were similar to those of wild type, see Fig.~\ref{fig:fits}B: sliding control cannot produce bend propagation, while curvature and normal force control are in good agreement with the experimental data. The parameters obtained from the fit of \textit{mbo2} beats are given in Table~\ref{tab:mbo2fits}.

\subsection*{Regulation of the beat by sliding} There are two related reasons why  the sliding control mechanism provides poor fits to the observed beating patterns. First, under sliding control the equations describing the flagellar beat become symmetric with respect to a change in sign of the arc length. As a consequence, the only possible solutions are standing waves. This can be most easily seen by using the plane-wave approximation, $\psi_1=\exp(-2\pi is/\lambda)$ with $\lambda$ real, for a symmetric beat  ($C_0=0$) regulated by sliding ($\beta=0$ and $\gamma=0$). In that case, Eq.~\ref{eq:asym} becomes an equation for the wave-length
\begin{align}
i\bar{\omega}=-(2\pi L/\lambda)^4-(2\pi L/\lambda)^2\bar{\chi}\quad,\label{eq:plane}
\end{align}
where { $\bar{\chi}=a_0^2L^2\chi/\kappa$} is the dimensionless complex sliding response coefficient. Note that this equation is symmetric with respect to the change $\lambda\to-\lambda$, and thus admits simultaneously waves traveling in both directions. In the absence of boundary asymmetries, these opposing waves interfere to form a standing wave. Thus to the extent that the asymmetry of the boundary can be neglected, the sliding control mechanism can not account for the bend propagation. 

\begin{figure}[!htb]
\begin{center}
\includegraphics{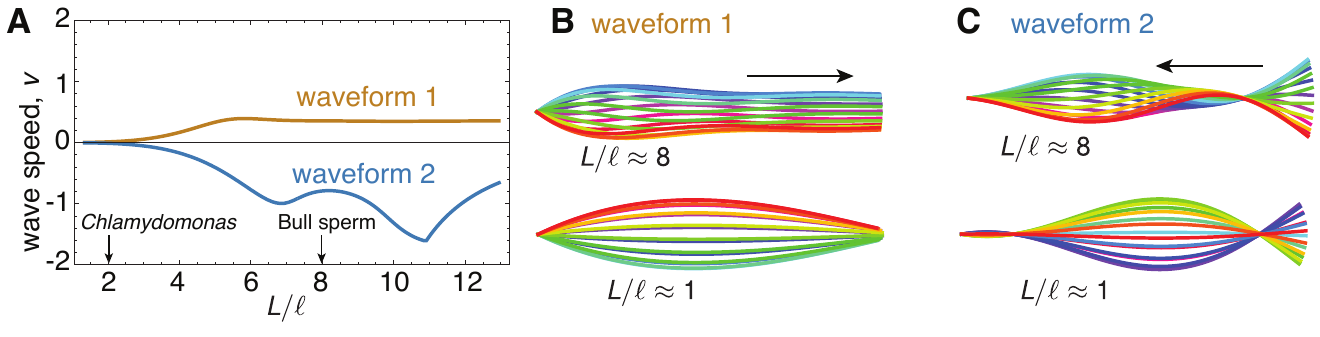}
\caption{{\bf The role of length in sliding regulated beats.} ({\bf A}) Wave speed versus relative length of the first two unstable waveforms of a freely swimming axoneme regulated by sliding. For short lengths, in the range of {\it Chlamydomonas}, the modes loose directionality and become standing waves. Long axonemes have directional waves that can travel either forward, as in waveform 1, or backwards, as in waveform 2. ({\bf B}) Two examples of waveform 1 for long (top) and short (bottom) axonemes. Note that the short axoneme shows a standing wave. ({\bf C}) beating patterns of waveform 2 for a long (top) and short (bottom) axonemes. In B and C arrows denote direction of wave propagation.
 \label{fig:slilen}}
\end{center}
\end{figure}

The second reason for the poor fits of the sliding control mechanism is that short flagella can not exhibit bend propagation, as already noted in \cite{camalet_generic_2000}. This is true even in the presence of a boundary asymmetry (from  the basal compliance). To understand this, we note that the dimensionless parameter $\bar{\omega}$ can be written as 
\begin{align}
\bar{\omega}=\left(\frac{L}{\ell}\right)^4\quad {\rm with}\quad \ell=\left(\frac{\kappa}{\xi_{\rm n}\omega}\right)^{1/4}\quad,
\end{align}
where $\ell$ is the characteristic length at which oscillations decay in a boundary driven axoneme \cite{machin_1958}. For the case of flagella short with respect to $\ell$, we can approximate $\bar{\omega}\approx0$ in Eq.~\ref{eq:plane}. The two allowed wavelengths are $\lambda_+=2\pi L/\sqrt{-\chi}$ and $\lambda_-=-\lambda_+$, where one can show that $\bar{\chi}=-\pi^2/4$ in order to satisfy the boundary condition of no torques on the distal end \cite{camalet_generic_2000}. These two modes correspond again to opposing waves, and imposing no torques on the basal end it can be shown that they must have equal amplitudes. The result is again a standing wave, irrespective of the basal asymmetry.

To verify this argument we studied the speed of bend propagation, which is defined as $v=\int_0^L|\psi_1|^2\partial_s\arg{\psi_1}\d s$, for two different flagellar lengths $L$, see  Fig.~\ref{fig:slilen}A. For axonemes with lengths $L$ similar or smaller than $\ell$, the resulting beats exhibit no bend propagation, Figs.~\ref{fig:slilen} B and C bottom. For axonemes significantly longer than $\ell$ wave propagation can be strong, Figs.~\ref{fig:slilen} B and C top. While in {bull sperm} we have $L/\ell\approx8$, which is enough to produce wave propagation \cite{riedelkruse_how_2007}; for {\it Chlamydomonas} we have $L/\ell\approx2$, which results in almost no bend propagation. Indeed, we estimated the wave speed of the sliding control waveform in Fig.~\ref{fig:fits} to be two orders of magnitude smaller than the observed value.

\subsection*{Regulation of the beat by normal forces}

Despite the good fits obtained for wild type and \textit{mbo2} axonemes, there are two observations that argue against normal force control. The first is that unlike the curvature response coefficient $\beta$, the mean value of the normal-force response coefficient $\gamma$ is much greater in mutants cells than in wild type cells (Table \ref{tab:mbo2fits}). The reason is that in \textit{mbo2} the small static curvature results in a small normal force, which requires a correspondingly larger response coefficient $\gamma$. In other words, to obtain agreement between the observed and theoretical waveforms,  the sensitivity of the motors to normal force needs to be much greater in the \textit{mbo2} mutant than in the wild type. Such a big difference in the response coefficient is not expected, given that the dynamic components of the mutant and wild type waveforms are very similar (compare Fig.~\ref{fig:tracked}C and \ref{fig:tracked}G).

\begin{figure}[!htb]
\begin{center}
\includegraphics{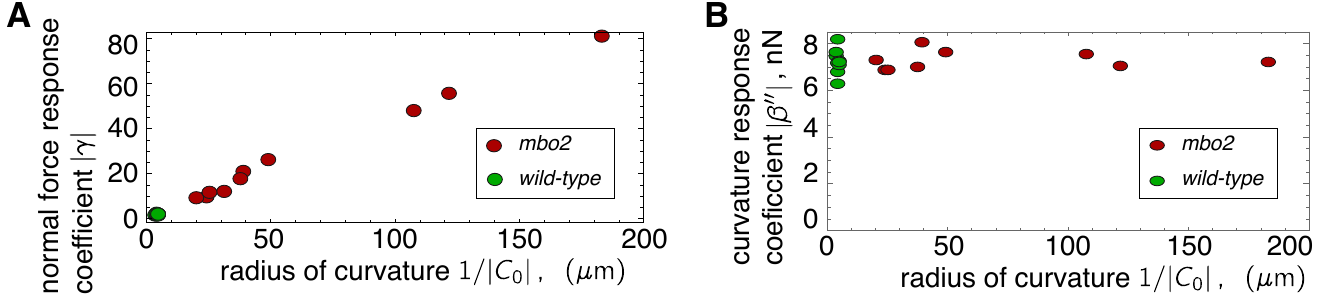}
\caption{{\bf Scaling of response coefficients with asymmetry.} ({\bf A}) Since the normal force $f_{\perp,1}$ is proportional to the static curvature $C_0$ but the fundamental mode stays unchanged, the normal force response coefficient $\gamma$ is inversely proportional to the curvature. In red values for {\it mbo2} and in green for {wild type}. ({\bf B}) The curvature control response coefficient $\beta''$ is  independent of the asymmetry, and stays constant even for a change in asymmetry of several orders of magnitude.
 \label{fig:normalasym}}
\end{center}
\end{figure}

The second argument against the normal force control is that $\gamma$ varies greatly from cell to cell for the case of \textit{mbo2}. This is due to the observation that, while the static curvature is small in \textit{mbo2} axonemes, it is variable. This leads to a large variability in $|\gamma|$ from axoneme to axoneme, which correlates strongly with the inverse of the static curvature (see Fig.~\ref{fig:normalasym}A), i.e.  $|\gamma|\propto |C_0|^{-1}$. Thus, even though the amplitude and phase of the first mode in \textit{mbo2} axonemes is very similar from axoneme to axoneme (Fig.~\ref{fig:tracked}B ($iii$) and ($iv$)), the response coefficients vary widely in amplitude. The reason for this correlation between $|\gamma|$ and $C_0$ is that, according to Eq.~\ref{eq:fndyn}, the dynamic component of the normal force is linearly proportional to the asymmetry $C_0$. This means that to preserve a similar fundamental dynamic mode, axonemes with a smaller static asymmetry $C_0$ require motors to have a higher response coefficient $|\gamma|$. By contrast, the curvature control response coefficient is highly consistent for wild type and mutant axonemes. 


In summary, despite the similarities in the dynamics of the beats of wild type and \textit{mbo2} axonemes, the normal force model requires very different values for the response coefficient $\gamma$ (and also the basal response coefficient $\chi_{\rm b}$) for wild type and \textit{mbo2} axonemes. Furthermore, the normal force model requires very large differences in $\gamma$ from axoneme to axoneme, despite the similarity in the dynamics between axonemes. Thus, we conclude that normal force is not a plausible parameter for controlling the ciliary beat.

\subsection*{Regulation of the beat by curvature}

The curvature control model provides a good fit to the experimental data for both {wild type} and {\it mbo2} axonemes (Figure~\ref{fig:fits}A and~B, middle panel). However, the strategy of curvature regulation  that we used is strikingly different from those previously studied \cite{mukundan2014motor,brokaw1971bend,riedelkruse_how_2007,brokaw_thinking_2009,morita2004effects}. While in previous work it was assumed that the static and dynamic response to curvature are of equal importance, here we found that when $\chi''$ and $\beta'$ were unconstrained, their best fit values were not significantly different from zero. We therefore set them both to zero (see Table~\ref{tab:wtfits}). We can understand this by writing Eq.~\ref{eq:asym} for symmetric plane waves regulated by curvature, which, after separating real and imaginary part, becomes
\begin{align}
\bar{\omega}+(2\pi L/\lambda)^2\bar{\chi}''-(2\pi L/\lambda)^3\bar{\beta}'&=0\nonumber\\
(2\pi L/\lambda)^4+(2\pi L/\lambda)^2\bar{\chi}'+(2\pi L/\lambda)^3\bar{\beta}''&=0\quad,\label{eq:ccplane}
\end{align}
where { $\bar{\beta}'=a_0L\beta'/\kappa$} is the dimensionless real curvature response coefficient,  $\bar{\beta}''=a_0L\beta''/\kappa$ is the complex one, and we assume  a passive response to sliding, $\chi'>0$ and $\chi''>0$. These equations show that  for a wave traveling forward ($\lambda>0$), $\beta'>0$ produces the active force to counter viscous effects of fluid and filament sliding, while $\beta''<0$ counters the elastic forces of filament bending and sliding. For the case in which the contribution of the response to sliding is smaller than that of the fluid and filament bending, we can divide the equations above to obtain $|\beta'/\beta''|\sim\bar{\omega}/(2\pi L/\lambda)^4$. Since for ${\it Chlamydomonas}$ $\bar{\omega}\sim2^4$ and $\lambda\sim L$, we find that $|\beta'/\beta''|\sim 10^{-2}$. In other words, for {\it Chlamydomonas}, elastic forces dominate over viscous forces, and thus $|\beta''|\gg|\beta'|$, as we obtained from the fits. Furthermore, in the limit of $\bar{\omega}=0$ (short lengths, low viscosity), we can set $\chi''=\beta'=0$ and still obtain plane waves. The opposite however is not true: for  $\beta''=0$ the balance of elastic forces can not be satisfied. The molecular implications of this finding will be expanded upon in the Discussion section.

\begin{figure}[!htb]
\begin{center}
\includegraphics{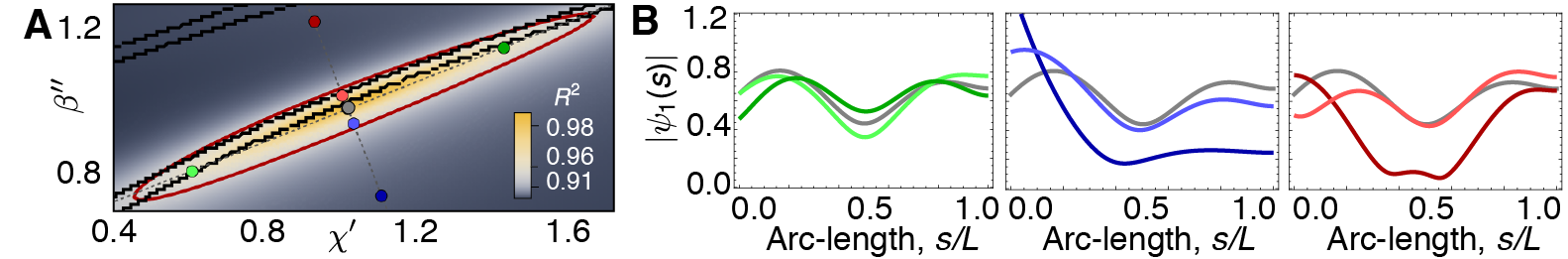}
\caption{{\bf Phase space of curvature control.} {({\bf A})} Heat map of the mean square distance  $R^2$ between the theoretical and a reference experimental beat as a function of the sliding response coefficient $\chi'$ and the curvature response coefficient $\beta''$. The ellipsoid delimits the region with $R^2=0.90$. Black lines delimit the region with a passive base. Moving along the long axis (green circles) affects the amplitude dip in the midpoint of the axonemem, see left panel in ({\bf B}). Moving along the small axis towards the region of active base results in waveforms with a large amplitude at the base (blue and red circles), see central and right panels in B. The axis in A are normalized by the reference fit, such that $(\chi'=1,\beta''=1)$ corresponds to the highest value of $R^2$. 
 \label{fig:curspace}}
\end{center}
\end{figure}

We therefore set $\chi'' = \beta'  = 0$ in the subsequent analysis. By doing so, we significantly simplify the model, because the number of free parameters is now only two,  $\chi'$ and $\beta''$. Note that the two parameters, $\chi_{\rm b}'$  and $\chi_{\rm b}''$, which characterize the stiffness and viscosity at the base respectively, are determined once $\chi'$ and $\beta''$ are specified as we are looking for oscillating solutions to the boundary value problem (see {\it Appendix}). Thus, the curvature control model is specified by just two free parameters, $\chi'$ and $\beta''$, which are specified by the sliding elasticity between doublet microtubules and the rate of change of axonemal curvature.

The average values of $\chi'$ and $\beta''$ varied little between wild type and \textit{mbo2} mutant axonemes (compare the third column of Table~\ref{tab:wtfits} with that of Table~\ref{tab:mbo2fits}). This accords with the observation that there is little difference in the dynamical properties of the beat between wild type and \textit{mbo2} axonemes. Furthermore, the standard deviation of $\chi'$ and $\beta''$ are small, indicating that there is little variation from axoneme to axoneme. Thus, the tight distribution of values of the parameters in the model reflects the similarity in the observed shapes in different axonemes. In other words, $\chi'$ and $\beta''$ are well constrained by the experimental data.

To understand what aspects of the experimental data specify these two parameters, we performed a sensitivity analysis on $\chi'$ and $\beta''$.
In Fig.~\ref{fig:curspace}A we show a density map of the mean square distance $R^2$ between the theoretical waveforms and a reference experimental beating pattern as a function of $\chi'$ and $\beta''$. A red ellipse delimits the region with $R^2>0.90$, which very closely coincides with the region where $\chi_{\rm b}'$ and $\chi_{\rm b}''$ are both positive, delimited by black lines. This is very important because negative values of the basal  parameters imply an active process at the base. Such an active process would drive a whiplike motion of the cilium as discussed in \cite{machin_wave_1958}. Evidently, the observed shapes of the beats rule out such a whiplike motion.

To investigate how the beat pattern is affected by variations in $\chi'$ and $\beta''$, as well as the existence of active processes in the base, we systematically varied $\chi'$ and $\beta''$ parallel and perpendicular to the long axis of the ellipse. Moving parallel affects the amplitude of the beat, with the middle-dip becoming more or less prominent, see green shapes in Fig.~\ref{fig:curspace}B. Moving perpendicular in the region of active base indeed results in whiplike beats, with a large amplitude at the base, see blue and red shapes in Figs.~\ref{fig:curspace}B. 

\begin{figure}[!ht]
\begin{center}
\includegraphics{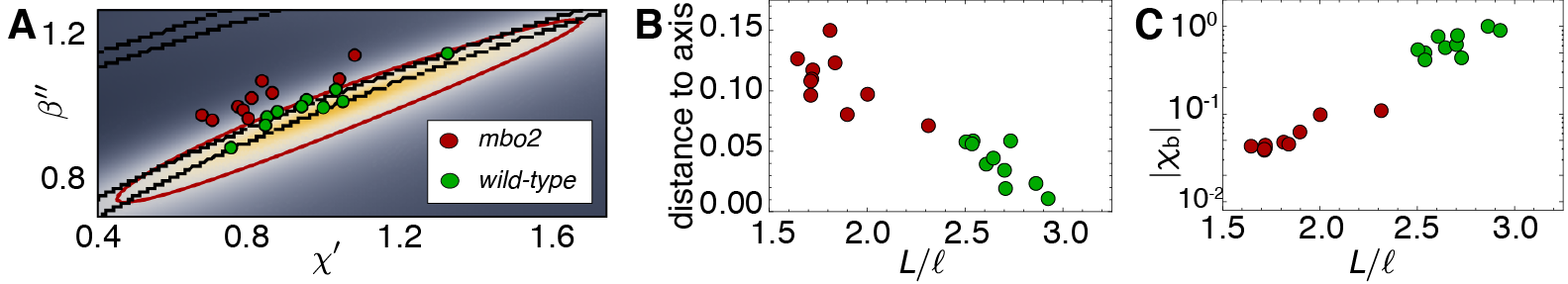}
\caption{{\bf Axonemal variability in phase space.} ({\bf A}) Circles represent values obtained from fits for each of the axonemes, green corresponds to {wild type} and red to {\it mbo2}. In the background we have the same heat map as in Fig.~\ref{fig:curspace}. Note that {\it mbo2} points lie away from {wild type} circles in the direction of the short axis of the ellipse. All values are normalized by those of the reference fit used also to normalize the heat map axis.  {({\bf B})} The distance  of the circles to the long axis of all fits shows a clear correlation with the axonemal length.  Note also that {\it mbo2} axonemes are systematically shorter than {wild type} axonemes. {({\bf C})} The basal response coefficient also correlates with the length, resulting in a high value for {wild type} axonemes, which are longer. The values are normalized by the value for a reference axoneme. 
 \label{fig:sensitivity}}
\end{center}
\end{figure}

To better understand the cell to cell variability we placed all the axonemes recorded in the $(\chi',\beta'')$  space, see Fig.~\ref{fig:sensitivity}A. Points scatter mainly along the long axis of the ellipse, where there is a large region of small shape variations. Importantly, we consistently see a shift perpendicular to the long axis between the {wild type} and {\it mbo2} mutant axonemes. This variation mainly comes from the difference in length between wild type and mutant axonemes, as can be seen in Fig.~\ref{fig:sensitivity}B. The implication of this length variation is a stiffening of the base, see Fig.~\ref{fig:sensitivity}C, which explains the variability in basal compliance between {\it mbo2} and {wild type} axonemes for curvature control in Tables~\ref{tab:wtfits} and \ref{tab:mbo2fits}.

\section*{Discussion}

In this work we imaged isolated axonemes of {\it Chlamydomonas} with high spatial and temporal resolution. We decomposed the beating patterns into Fourier modes and compared the fundamental mode, which is the dominant dynamic mode, with  theoretical predictions of the three motor control mechanisms illustrated in Fig.~\ref{fig:axoscheme}.  The sliding control model provided a poor fit to the experimental data. We argued that the reason for this is that sliding control cannot produce wave propagation for axonemes as short as those of {\it Chlamydomonas}, see Fig.~\ref{fig:slilen}. While the normal force model provided good fits to the experimental data, it relies on the presence of static asymmetry \cite{bayly2015analysis,mukundan2014motor}, which varies greatly between the {\it mbo2} and {wild type } axonemes. As a result of this large difference in static curvature, the control parameters in this model had to be varied over a large range to fit the data from the different axonemes, see Fig.~\ref{fig:normalasym}. Because the waveforms of {\it mbo2} and wild type axonemes have very similar dynamic characteristics, such variation in the control parameter seems implausible. Finally, the curvature control model provided  a good fit to the experimental data with similar parameters for {\it mbo2} and wild type axonemes. Thus, we conclude that only the two-dimensional curvature-control model is fully consistent with our experimental data. 

A potential caveat of the {model} used here is that it is two-dimensional. Importantly, in order to simplify the geometry, the model only contains one pair of doublet microtubules. In the three-dimensional axoneme, there are pairs of doublets on opposite sides of the axoneme which, due to the approximate rotational symmetry of the structure, are bent in opposite directions when their associated dyneins are activated. A key feature of dynein coordination is that motors on either side of the axoneme are antagonistic (i.e. in a ``tug-of-war''), such that when dyneins on either side are active they bend the axoneme in opposite directions \cite{satir1989splitting}. Both sliding control and curvature control ensure that the tug-of-war is unstable such that if dyneins on one side begin to dominate, then they completely dominate in a ``winner-takes-all'' scenario. To capture this  idea of reciprocal inhibition by opposing dyneins in the two-dimensional model, dyneins are anchored with opposite orientations to each of the doublets in the pair \cite{camalet2000generic}. While this captures the essential features of the sliding control and curvature control models, it oversimplifies the normal force model, because in the three-dimensional axoneme there are radial and transverse forces acting on the doublets as the axoneme bends. Yet the two-dimensional model does not distinguish between them. To bridge this gap, in other work \cite{3dpaper} we use a full three-dimensional description of the axoneme to calculate the radial and transverse stresses. The three-dimensional model shows that even when there is a static curvature (without twist), normal (transverse) forces are not antagonistic across the centerline and therefore cannot serve as a control parameter for motors.


\subsection*{Relation with past work}
Earlier results showed that sliding control can account for the beating patterns of sperm \cite{riedelkruse_how_2007,brokaw2009thinking,brokaw_molecular_1975}. This result is not inconsistent with our results because the bull sperm axoneme is approximately five times longer than the {\it Chlamydomonas} axoneme and we have shown that sliding control can work for long axonemes, while for short axonemes it produces no bend propagation, see also \cite{brokaw2005computer}. Thus, it is possible that different control mechanisms operate in different cilia and flagella, with sliding control being used in longer axonemes and curvature control being used in shorter ones. However we do note that curvature control models can  account for the bull sperm data \cite{riedelkruse_how_2007} as well as  data from other sperm \cite{brokaw2002computer,brokaw1985computer,bayly2015analysis}, so there is no strong morphological evidence favoring either sliding or curvature control in sperm.  Other studies have shown that the normal force model gives beating patterns that resemble those of sperm  \cite{lindemann_model_1994, bayly2015analysis, bayly2014equations }. These models, like ours, rely on there being an asymmetry. What we have shown is that the control parameter depends critically on this asymmetry while the similarity in the fundamental dynamic modes of {\it mbo2} and wild type suggests that the parameters should be similar. Thus, the curvature control model, unlike the other two models, robustly describes symmetric and asymmetric beats in  short and long axonemes, and could serve as a ``universal'' regulator of flagellar mechanics.

\subsection*{Dynamic curvature control as a mechanism for motor regulation}
One interesting feature of our curvature control model is that in order to describe the observed beating patterns the motor force depends only on the time derivative of the curvature. This follows from the fact that the curvature response function $\beta(\omega)$ has no real part, see tables, and thus vanishes at zero frequency, $\beta(\omega=0)=0$. Such a model is fundamentally different from the current views of curvature control, in which motors are thought to respond to curvature \cite{brokaw_bend_1971, brokaw2002computer,brokaw_thinking_2009, machin_wave_1958} and not to its time derivative.  While motors can respond to time derivatives of sliding displacement through their force-velocity relation, it is hard to understand how a similar mechanism could apply to curvature. A more plausible mechanism giving rise to a response to the time derivative of curvature is an adaptation system analogous to that of sensory systems, like the signaling pathway of bacterial chemotaxis \cite{macnab1972gradient, koshland1982amplification, yi2000robust, shimizu2010modular}. In an adaptation mechanism motor activity would be ``remembered'', and the average activity over the past times  would in turn down-regulate the activity of the motor on a long time-scale. Such regulation could occur, for example, via phosphorylation sites in the dynein regulatory complex or the radial spokes \cite{witman2009chlamydomonas,smith2004radial,porter20009,lindemann2003structural}. Just as methylation of the chemoreceptors of bacteria modifies their ligand affinity, phosphorylation of regulatory elements within the axoneme could modify the motor sensitivity to curvature.

\begin{figure}
\begin{center}
\includegraphics{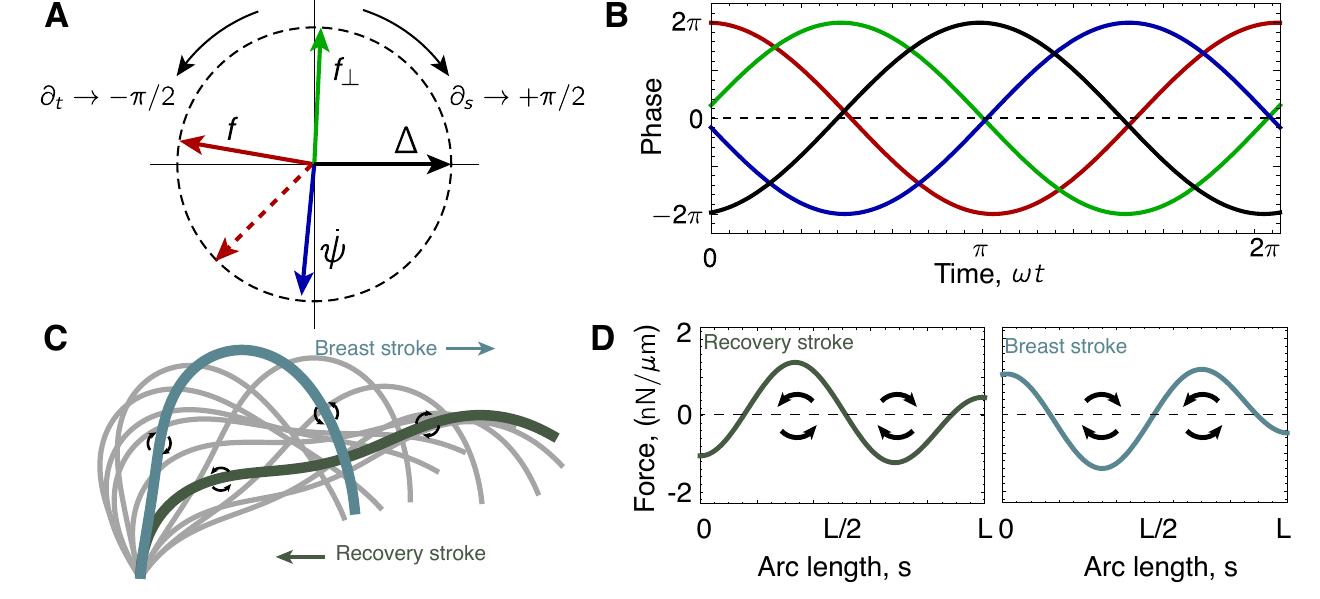}
\caption{{\bf Phase delays and active force distribution on the axoneme.} {({\bf A})} Polar representation of the phase of sliding $\Delta$,  curvature $\dot{\psi}$, normal force $f_\perp$ and the sliding force $f$ for a beating axoneme regulated by curvature. { For a plane wave  we have $\psi_1=\exp{i(\omega t-2\pi s/\lambda)}$. A time derivative thus adds a counter-clockwise $\pi/2$ phase, while a spatial derivative adds $\pi/2$ clockwise. Deviations from this come from the solution deviating from a plane wave due to the boundary conditions. In dashed red we show Machin's prediction for an optimum flagellum.}  {({\bf B})} Time evolution of the quantities in A. {({\bf C})} Beating axoneme with two shapes highlighted during the breast and recovery strokes. The black arrows represent direction of motion, and the circular arrows represent the local sliding force. ({\bf D}) Sliding force density over arc length for the two shapes highlighted in C.  \label{fig:phase}}
\end{center}
\end{figure}

\subsection*{Independence of static and dynamic waveform components}
Our dynamic curvature control model adds to the view that dynamic and static components of the beat are regulated independently. The problem with models in which dynein activity is regulated by the instantaneous value of the curvature \cite{mukundan2014motor, brokaw_bend_1971, brokaw_thinking_2009, machin_wave_1958} is that both the static and dynamic components of the beat would contribute to regulation and hence the dynamic component of the waveform would be highly dependent on the static component \cite{thesis}. This potential problem was noted by Eshel and Brokaw in \cite{eshel1987new}. Our dynamic curvature control model provides a solution to this problem because static curvature is ``adapted'' away. We now bring together several lines of evidence supporting the notion that the static and dynamic modes are separable in their origin and in their affect on the beat
\begin{enumerate}
 \item The waveform of {\it Chlamydomonas} can be mathematically decomposed into a bending wave superimposed on a static asymmetry (a circular arc), see \cite{eshel1987new,geyersub}. 
 
\item Dynamic and static components of the beat can exist independently of each other. This is evidenced by the existence of bent non-motile cilia at low ATP concentrations on the one hand, as well as symmetrically beating mutants on the other \cite{geyersub}.
 
\item The waveform of {\it mbo2} has  a similar fundamental dynamic mode as that of wild type, compare Fig.~\ref{fig:tracked} A $(iii)$ and B $(iii)$. However, the static mode is absent in the former, compare Fig.~\ref{fig:tracked} A $(ii)$ and B $(ii)$. The same also holds for the  two beating modes of the uniflagellar mutant \cite{eshel1987new}. Thus, altering the static mode of {\it Chlamydomonas} has little effect on the dynamic mode.

\item The dynamic motor response coefficients are largely independent of the asymmetry, and very similar for {\it mbo2} and {wild type} axonemes, see Fig.~\ref{fig:sensitivity} A.
\end{enumerate}

If the dynamic and static modes are indeed independently controlled, the dynamic motor response is robust to changes in the asymmetry. This has important biological implications: power generation (the beat) and steering (the asymmetry) can be independently controlled so that the swimming direction can be adjusted without having to alter the motor properties. 


\subsection*{Phase relations during the beat}
The dynamics of the axoneme can be  simplified using a phase plot, see Fig.~\ref{fig:phase}A. In this diagram, each arrow represents the normalized complex value of the corresponding quantity at the midpoint of the axoneme ($s=L/2$). Over time, the arrows rotate counter-clockwise at a homogeneous speed, preserving their phase relations. In the plane wave limit, an arc length derivative adds a delay of $\pi/2$, which is why the curvature lags almost one quarter behind the sliding. The motor force is itself delayed one quarter of a cycle with respect to curvature, which makes it  in anti-phase with the sliding. Finally, since $C_0<0$, from Eq.~\ref{eq:fndyn} we see that the normal force is in anti-phase with the curvature, see Fig.~\ref{fig:phase}B for a time trace. 

Importantly these phase relations deviate from the prediction made by Machin in \cite{machin_wave_1958} for the most energy efficient flagellum  (note that his bending moment $B$ is $\sim -F$, and his displacement $y$ is $\sim \int\Delta\d s$). In fact, using the same plane wave arguments as in \cite{machin_wave_1958}, we already showed after Eq.~\ref{eq:ccplane} that the elastic contribution to the sliding response dominates the viscous component by almost two orders of magnitude, which results in the one quarter delay with respect to curvature. Thus, the {\it Chlamydomonas} flagella is not optimal under Machin's assumptions.





\section*{Appendices}
\subsection*{Non-linear dynamics of the axoneme}
The equations that describe the dynamics of the axoneme are obtained by balancing mechanical and fluid forces. We proceed using a variational approach similar to that in \cite{camalet_generic_2000, riedelkruse_how_2007,mukundan2014motor}, and introduce the work functional $U$ given by 
\begin{align}
U&= \int_0^L\left[\frac{\kappa}{2}\dot{\psi}^2  +f\Delta+ f_\perp(a-a_0) +\frac\Lambda2({ \dot{\bf r}}^2-1)\right]\d s+\frac{k_{\rm b}}{2}\Delta_{\rm b}^2\quad,
\end{align}
where $\kappa$ is the bending rigidity, $f\Delta$ the work performed by the motors, and $k_{\rm b}$ the stiffness of cross-linkers at the base. The normal force $f_\perp$ is a Lagrange multiplier that ensures that $a=a_0$, see \cite{mukundan2014motor,bayly2015analysis,thesis} for the more general case of variable $a$. Similarly, $\Lambda$ is a multiplier that ensures the incompressibility constraint $\dot{\bf r}^2=1$, and is related to the tension through { $\tau=\Lambda+\kappa\dot{\psi}^2-a_0F\dot{\psi}$}, where $F=-\int_s^Lf(s')\ds'$ \cite{camalet_generic_2000}.

The mechanical force that the axoneme exerts on the fluid is given by $\delta U/\delta {\bf r}$, and calculating it requires computing $\delta \dot{\psi}$. From the relation ${\bf r}(s)={\bf r}_0+\int_0^s(\cos(\psi(s')),\sin(\psi(s')))\d s'$, where ${\bf r}_0$ is the position of the base, it follows that $\dot{\psi}={\bf n}\cdotp\dot{\bf t}$ and $\delta\dot\psi={\bf n}\cdotp\delta\ddot{\bf r}$. Using this, we arrive at $\delta U/\delta{\bf r}=\partial_s\left[(\kappa\ddot{\psi}-a_0f){\bf n}-\tau{\bf t}\right]$ \cite{camalet_generic_2000,thesis}. Similarly, the net sliding force exerted at the base is $\delta U/\delta \Delta_{\rm b}=-F(0)+k_{\rm b}\Delta_{\rm b}$ \cite{mukundan2014motor}. To obtain the dynamics of the axoneme we balance these mechanical forces by the fluid friction $\hat{\bf \Pi}\cdotp\dt{\bf r}$ and the basal friction $\xi_{\rm b}\dt \Delta_{\rm b}$, which results in 
\begin{align}
\partial_t{\bf r}&=-(\xi_{\rm n}^{-1}{\bf n}{\bf n}+\xi_{\rm t}^{-1}{\bf t}{\bf t})\cdotp\partial_s\left[ (\kappa\ddot{\psi} -a_0 f){\bf n}-\tau{\bf t} \right]\\
\partial_t\Delta_{\rm b}&=-\xi_{\rm b}^{-1}( k_{\rm b}\Delta_{\rm b}-F(0) )\label{eq:D0nonlin}\quad,
\end{align}
We can also calculate a dynamic equation for the tangent angle using that $\partial_t\dot{\bf r}={\bf n}\partial_t\psi$, which results in
\begin{align}
\label{eq:angnonlin}
\partial_t\psi&=\xi_{\rm n}^{-1}(-\kappa\ddddot{\psi} +a_0 \ddot{f} +   \dot{\psi}\dot{\tau} + \tau\ddot{\psi}) +\xi_{\rm t}^{-1} \dot{\psi}(\kappa\dot{\psi}\ddot{\psi}-a_0f\dot{\psi}+\dot{\tau})\quad.
\end{align}
This equation contains no information about the trajectory of the basal point ${\bf r}_0(t)$, which can be determined from the condition that the total force on the cilium vanishes \cite{friedrich2010high,thesis,johnson1979flagellar}.

The tension $\tau$ and normal force $f_\perp$ are obtained by imposing the corresponding constraints. For the case of the tension we take the time derivative of $\dot{\bf r}^2=1$. This gives  ${\bf t}\cdotp\partial_t\dot{\bf r}=0$, where we can replace the dynamic equation for ${\bf r}$. For the normal force $f_\perp$ we use the force balance $\delta G/\delta a=0$ \cite{camalet_generic_2000,mukundan2014motor}. The resulting constraint equations are
\begin{align}
\frac{\xi_{\rm n}}{\xi_{\rm t}}\ddot{\tau}- \dot{\psi}^2\tau&=-\dot{\psi}(\kappa\dddot{\psi}-a_0\dot{f}) + \frac{\xi_{\rm n}}{\xi_{\rm t}}\partial_s[ \dot{\psi}_0 (a_0f-\kappa\ddot{\psi})]\nonumber\\
f_\perp&=F\dot{\psi}\quad.\label{eq:fnnonlin}
\end{align}
Finally, to completely characterize the dynamic equations we need to use boundary conditions. These represent force and torque balances at the ends of the filament pair, and are obtained from the boundary terms of the variational calculation. For the case of free ends considered in this work we have 
\begin{align}
\label{eq:bctime}
{s=0:}\quad  \kappa\ddot{\psi}(0)= a_0f (0)\;\;\; ,\;\;\; \tau(0)&=0\;\;\; ,\;\;\;\kappa\dot{\psi} (0)=a_0F_{\rm b} \nonumber\\
{s=L:}\quad  \kappa\ddot{\psi}(L)= a_0f (L)\;\;\; ,\;\;\; \tau(L)&=0\;\;\; ,\;\;\;\dot{\psi} (L)=0 \quad,
\end{align}
where the basal force is $F_{\rm b}=k_{\rm b}\Delta_{\rm b}+\xi_{\rm b}\partial_t\Delta_{\rm b}$. Together with a suitable motor model that provides the dynamics of the sliding force (see Eq.~\ref{eq:fullmotmod}), the equations above allow to compute the state of the cilium over time.

\subsection*{Sign convention}
{ It is convenient to clarify the sign convention used in this paper for the geometry and the forces. The tangent angle is measured with respect to the horizontal ${\bf x}$ axis and grows counter-clockwise (the ${\bf xy}$ frame has the usual orientation, see Fig.~\ref{fig:axoscheme}A ). With this choice the mid-point of the shape shown in Fig.~\ref{fig:tracking}A $(i)$ and $(ii)$, in black, has a negative angle. Since the flagellum swims counterclockwise, the tangent angle slowly grows positive over time, Fig.~\ref{fig:tracking}A $(iii)$. This applies to all flagella used in this work, which were imaged using an inverted microscope such that axonemes were seen ``from behind''. With this convention the mean angle has a negative slope, as shown in Fig.~\ref{fig:tracked}A $(ii)$, which corresponds to $C_0<0$. From Eq.~\ref{eq:sliding} we then have $\Delta<0$ for the simple case in which $\Delta_{\rm b}=0$. A shape with $C_0<0$ and  $\Delta_{\rm b}=0$ requires that the bottom filament stands out at the tip, which fixes the sliding sign: sliding is positive if the bottom filament slides towards the distal end. Thus basal and distal sliding in Fig.~\ref{fig:axoscheme}A are negative. 

Because dynein is a minus end directed motor, the sliding force density $f$ is taken to be positive when the sliding is negative. This corresponds to a motor with its head (green circle in Fig.~\ref{fig:axoscheme}A) in the bottom filament moving towards the base. In other words, the force is positive when the bottom filament is tensed towards the distal end and the top one towards the basal end. According to Eq.~\ref{eq:D0nonlin} a positive static force $f>0$ is opposed by a basal force $F_{\rm b}<0$, which results in negative sliding. Such a positive force, like the one given by Eq.~\ref{eq:statfor} for $C_0<0$, corresponds to a negative value of the integrated sliding force $F$. The static force balance $\kappa\dot{\psi}_0=a_0 F_0$ establishes then that a negative integrated force results in a negative curvature, as is the case for the {\it Chlamydomonas} axoneme. 
}
\subsection*{Asymmetric equation for the fundamental mode}
The periodic dynamics of the tangent angle can be decomposed in Fourier modes as indicated in Eq.~\ref{eq:modes}. For asymmetric beating patterns in which $\dot{\psi}_0\neq0$, the static mode is characterized by the force balance $\kappa\dot{\psi}_0=a_0 F_0$, obtained from integrating Eq.~\ref{eq:angnonlin} and using the boundary conditions. While the static component of the tension vanishes, the normal force has a static contribution given by $f_{\perp,0}=\kappa\dot{\psi}_0^2/a_0$.  The dynamics of a small amplitude oscillation dominated by the fundamental mode can be described by expanding Eqs.~\ref{eq:angnonlin} and \ref{eq:fnnonlin} around the static component. This results in
\begin{align}
i{\omega}\xi_{\rm n}\psi_1&=-\ddddot{\psi_1} + a_0\ddot{f}_1+\dot{\psi}_0 \dot{\tau}_1 + \ddot{\psi}_0\tau_1+\frac{\xi_{\rm n}}{\xi_{\rm t}}\dot{\psi}_0(\kappa\dot{\psi}_0\ddot{\psi}_1 - a_0\dot{\psi}_0 f_1+\dot{\tau}_1)\quad,\nonumber\\
\frac{\xi_{\rm n}}{\xi_{\rm t}}\ddot{\tau}_1- \dot{\psi}_0^2\tau_1&=-\dot{\psi}_0(\kappa\dddot{\psi}_1-a_0\dot{f}_1) + \frac{\xi_{\rm n}}{\xi_{\rm t}}\partial_s[ \dot{\psi}_0 (a_0f_1-\kappa\ddot{\psi}_1)]\quad ,
\label{eq:dynforbal}
\end{align}
where non-linear terms in the fundamental mode have been neglected. This pair of equations is the generalization of the equations for the symmetric beat \cite{machin_wave_1958,camalet1999self}. In them, the fundamental mode is coupled to the static mode. Expanding the expression of the normal force in Eq.~\ref{eq:fnnonlin}, we obtain that its fundamental mode is given by $f_{\perp,1}=\dot{\psi}_0(F_1+ \kappa \dot{\psi}_1/a_0)$.

The equations above can be rendered into dimensionless form using the following rescalings: $\bar{s}=s/L$, $\bar{\Delta}=\Delta/a_0$, $\bar{f}=a_0L^2f/\kappa$, $\bar{\tau}=L^2\tau/\kappa$, $\bar{k}_{\rm b}=a_0^2L k_{\rm b}/\kappa$ and $\bar{\xi}_{\rm b}=a_0^2L \omega \xi_{\rm b}/\kappa$. This choice results in the additional rescalings $\bar{f}_\perp=a_0L^2f_\perp/\kappa$, $\bar{\chi}=a_0^2L^2\chi/\kappa$, $\bar{\beta}=a_0L\beta/\kappa$ and $\bar{\gamma}=\gamma$, since $\gamma$ is already dimensionless. For the particular case in which the static shape has constant curvature $\dot{\psi}_0=C_0$, Eqs.~\ref{eq:dynforbal} then reduce to the asymmetric beat equations used in the main text.

Eqs.~\ref{eq:asym} together with Eq.~\ref{eq:fullmotmod} form a system of ordinary differential equations. Using boundary conditions the discrete spectrum of solutions can be obtained, see \cite{cross1993pattern, camalet_generic_2000,thesis}. While the system is of sixth order, it contains an integral term in the expression for the normal force, Eq.~\ref{eq:fndyn}. It is thus convenient to convert the system to seventh order by taking the derivative of Eq.~\ref{eq:fndyn}, which eliminates the integral term. Provided values for the response coefficients $\chi$, $\beta$ and $\gamma$ we can then use the {\it ansatz} $\psi_1=A\e^{ks}$ to obtain a characteristic polynomial of order seven in $k$. The general solution to the boundary value problem is $\psi_1= \sum_{i=1}^{7} A_i\e^{k_is}$, where the roots $k_i(\chi,\beta,\gamma)$ of the characteristic polynomial are implicit functions of the motor response coefficients. The amplitudes $A_i$  are then determined, up to an arbitrary factor, imposing that the boundary conditions be satisfied. Determining the amplitudes will in turn result in a fixed discrete spectrum of solutions for the possible basal compliances. Conversely, if the basal compliance is provided, calculating the amplitudes will return a discrete set of solutions for the real and imaginary parts of one of the response coefficients. This discrete set are the critical modes in \cite{camalet_generic_2000}, of which here we have shown two examples in Fig.~\ref{fig:slilen}.

\subsection*{Fitting procedure}
The fitting procedure was done as follows. Given a set of values for the response coefficients $\chi$, $\beta$ and $\gamma$, a solution $\psi_{\rm the}$ was obtained in the manner described in the previous section, up to an arbitrary complex amplitude. Given this solution, the force balance $\chi_{\rm b}\Delta_{\rm b}=F(0)$ allows to determine $\chi_{\rm b}$. If the value for the real or imaginary parts of $\chi_{\rm b}$ were negative, corresponding to an active base, the solution was discarded. If they were positive, then the complex amplitude was chosen as to minimize the mean square displacement $R^2(\psi_{\rm the},\psi_{\rm exp})$ given by 
\begin{align}
R^2=1-\frac{\sum_{i=1}^N|\psi_{\rm exp}(s_i)-\psi_{\rm the}(s_i)|^2}{\sum_{i=1}^N|\psi_{\rm exp}(s_i)|}\quad,
\end{align}
where $N=20$ and the points $s_i$ were equally spaced along the axonemal length. Finally, a value of $R^2<1$ was obtained. This $Q$ function, which takes as input response coefficients, was maximized with the routine FindMinimum of Mathematica 10 using the Principal Axis method.

\subsection*{Estimation of mechanical parameters of the axoneme}
The only parameters entering the problem are the effective stiffness $\kappa$ and spacing $a_0$ of the opposing filament description, and the two friction coefficients $\xi_{\rm n}$ and $\xi_{\rm t}$. For each motor model the motor response was used to best fit the data, and the basal compliance was obtained by solving the problem as described above.

A doublet has $24$ protofilaments compared to $13$ in a microtubule. We thus estimate the bending stiffness to be  $\kappa_{\rm db}\approx 2\kappa_{\rm mt}\approx46\,{\rm pN}\,\um^2$, with $\kappa_{{\rm mt}}\approx23\,{\rm pN}\,\um^2$ the microtubule stiffness measured in \cite{gittes_flexural_1993}. If we consider for simplicity an additive effect among the sliding doublets of the axoneme, we have  ${\kappa}=9\kappa_{{\rm db}}\approx400\,{\rm pN}\,\um^2$,  comparable to measurements of { sea urchin sperm} \cite{howard}.

The radius of  the axoneme is approximately $0.2\,\um$ \cite{nicastro2006molecular}. Considering that three doublets can be simultaneously subjected to significant active sliding during the beat we estimate $a_0\approx0.06\,\um$.

The friction coefficients are $\xi_{\rm t}\approx2\pi\mu/(\ln(2L/a_0)-1/2)$ and $\xi_{\rm n}\approx2\xi_{\rm t}$, where $\mu$ is the viscosity \cite{johnson1979flagellar, gray1955propulsion, cox1970motion}. For water  at $22\,^{\rm o}{\rm C}$ we have $\mu=0.96\, 10^{-3}{\rm pN\,s}\,\mu{\rm m}^{-2}$, which for $L\approx\,10\,\mu{\rm m}$ results in $\xi_{\rm t}\approx0.0017\,{\rm pN}\,{\rm s}\,\um^{-2}$ and  $\xi_{\rm n}\approx0.0034\,{\rm pN}\,{\rm s}\,\um^{-2}$.

\nolinenumbers


\end{document}